\documentclass[usenatbib]{mn2e}
\usepackage{natbibmnfix,graphicx,times}

\newcommand{\kel}{\mbox{ K}}

\newcommand{\Mpc}{\mbox{ Mpc}}

\newcommand{\yr}{\mbox{ yr}}

\newcommand{\hunits}{\mbox{ km s$^{-1}$ Mpc$^{-1}$}}

\newcommand{\bq}{\begin{equation}}
\newcommand{\eq}{\end{equation}}
\newcommand{\bqa}{\begin{eqnarray}}
\newcommand{\eqa}{\end{eqnarray}}
\newcommand{\deriv}{{\rm d}}

\newcommand{\fcoll}{f_{\rm coll}}
\newcommand{\fcollst}{f_{\rm coll,ST}}
\newcommand{\mmin}{m_{\rm min}}
\newcommand{\bxio}{\bar{x}_i}
\newcommand{\bxh}{\bar{x}_{\rm HI}}

\newcommand{\hii}{HII }
\newcommand{\lya}{Ly$\alpha$ }
\newcommand{\lyans}{Ly$\alpha$}

\newcommand{\apj}{ApJ}
\newcommand{\apjl}{ApJ}
\newcommand{\apjs}{ApJS}
\newcommand{\aap}{A \& A}
\newcommand{\aj}{AJ}
\newcommand{\mnras}{MNRAS}
\newcommand{\physrep}{Physics Reports}
\newcommand{\prd}{Phys Rev D}
\newcommand{\nat}{Nature}

\title[Characteristic scales during reionization]{Characteristic scales during reionization}

\author[Furlanetto et al.]{Steven R. Furlanetto,$^1$\thanks{Email:sfurlane@tapir.caltech.edu} Matthew McQuinn,$^2$ and Lars Hernquist$^2$\\ 
$^1$Division of Physics, Mathematics, \& Astronomy; California Institute of Technology; Mail Code 130-33; Pasadena, CA 91125, USA \\ 
$^2$Harvard-Smithsonian Center for Astrophysics, 60 Garden St., Cambridge, MA 02138, USA} 

\begin{document}

\maketitle

\begin{abstract}
One of the key observables of the reionization era is the distribution of neutral and ionized gas.  Recently, Furlanetto, Zaldarriaga, \& Hernquist developed a simple analytic model to describe the growth of \hii regions during this era.  Here, we examine some of the fundamental simplifying assumptions behind this model and generalise it in several important ways.  The model predicts that the ionized regions attain a well-defined characteristic size $R_c$ that ranges from $\sim 1 \Mpc$ in the early phases to $\ga 10 \Mpc$ in the late phases.  We show that $R_c$ is determined primarily by the bias of the galaxies driving reionization; hence measurements of this scale constrain a fundamental property of the first galaxies.  The variance around $R_c$, on the other hand, is determined primarily by the underlying matter power spectrum.  We then show that increasing the ionizing efficiency of massive galaxies shifts $R_c$ to significantly larger scales and decreases the importance of recombinations. These differences can be observed with forthcoming redshifted 21 cm surveys (increasing the brightness temperature fluctuations by up to a factor of two on large scales) and with measurements of small-scale anisotropies in the cosmic microwave background.  Finally, we show that stochastic fluctuations in the galaxy population only broaden the bubble size distribution significantly if massive galaxies are responsible for most of the ionizing photons. We argue that the key results of this model are robust to many of our uncertainties about the reionization process.
\end{abstract}


\section{Introduction}
\label{intro}

The reionization of hydrogen in the intergalactic medium (IGM) is the hallmark event of the first generation of luminous objects, marking the time at which they affected every baryon in the Universe.  In the past few years, astronomers have made enormous strides in understanding this transition, but the emerging picture is complex.  The rapid onset of \citet{gunn65} absorption in the spectra of $z>6$ quasars \citep{becker01,fan02,white03} indicates that reionization probably ended at $z \sim 6$ (albeit with large variance among different lines of sight: Sokasian et al. 2003 [their Fig. 10]; Wyithe \& Loeb 2004a; Oh \& Furlanetto 2005).  On the other hand, the detection of a large optical depth to electron scattering for cosmic microwave background (CMB) photons \citep{kogut03} indicates that the process began at $z \ga 15$ (although only with relatively low confidence).  Reionization thus appears to be a complicated process, a conclusion strengthened by several other studies \citep{theuns02-reion,wyithe04-prox,mesinger04,malhotra04}.

For these reasons, untangling the puzzle of how the ``twilight zone"
of reionization ended is one of the most exciting areas of cosmology.
A number of new techniques are being developed to study this epoch.
Perhaps the most interesting focus not on \emph{when} reionization
occurred but on \emph{how} it happened: the evolution of the
distribution of ionized and neutral gas can teach us an enormous
amount about how the ionizing sources interacted with the IGM.  The
most promising approach is to observe the redshifted 21 cm transition
of neutral hydrogen in the IGM, which would allow us to perform
``tomography" of the entire reionization process, watching in detail
as the \hii regions grew and filled the universe (see
\citealt{furl04-skarev} for a recent review).  If the technical
challenges inherent to these observations (including terrestrial radio
interference, ionospheric distortions, and the Galactic and
extragalactic foregrounds) can be overcome, such experiments promise
to reveal the entire history of reionization.  A number of other
techniques will complement these studies, including the kinetic
Sunyaev-Zel'dovich (kSZ) effect from patchy reionization in CMB maps
\citep{aghanim96, gruzinov98,knox98}, absorption spectra of
high-redshift quasars, and the spatial distribution of \lyans-selected
galaxies \citep{furl04-lya}.

The key observables in each of these techniques are the sizes of \hii regions and their distribution throughout the universe.  These bubbles presumably appear around the first protogalaxies; as more ionizing sources form the bubbles grow and merge, eventually filling all of space.  A consensus is now emerging that these bubbles likely attain quite large ($\ga 10$ comoving Mpc) scales throughout most of reionization:  analytic arguments suggest large-scale variations in the ionizing emissivity \citep{barkana04-fluc}, and numerical simulations show that indeed the \hii regions rapidly grow to sizes comparable to the simulation boxes \citep{gnedin00,razoumov02,ciardi03-sim,sokasian03,sokasian04}.  \citet{wyithe04-var} also argued that the ionized bubbles must achieve sizes $\ga 10$ physical Mpc at the end of reionization, and \citet{cen05} showed that, if dwarf galaxies in present-day clusters were responsible for reionization, the \hii regions must reach similar sizes.  Measuring the sizes of these bubbles throughout reionization, as well as their locations relative to galaxies and quasars, can teach us a great deal about structure formation in the early universe.

Recently, \citet[hereafter FZH04]{furl04-bub} constructed the first quantitative model for the distribution of bubble sizes throughout reionization.  From such a model, a wide range of predictions are possible, including those for 21 cm surveys \citep{furl04-21cmtop}, quasar absorption spectra \citep{furl04-lya}, the kSZ effect \citep{zahn05,mcquinn05}, and \lya galaxy surveys \citep{furl04-lya,furl05-lyagal}.  The model is based on the excursion set formalism familiar from the dark matter halo mass function \citep{bond91}.  It uses a simple photon counting argument to build \hii regions around proto-clusters of galaxies.  We review the basic model in \S \ref{fzh}.  One key prediction for all the observables is that the bubbles attain a well-defined characteristic size.  In \S \ref{charsize} and \ref{width} we study how this scale depends on the underlying source population and the cosmology.  

To construct an analytic model, FZH04 made a number of simplifying
assumptions.  These include spherical symmetry, a constant ionizing
efficiency throughout the universe, a simple approach to the halo mass
function, an approximate treatment of the inhomogeneous IGM, and
ignoring ``shot noise" in the galaxy distribution as well as the
potentially bursty nature of star formation at high redshifts.
While some of these factors (such as the IGM and spherical symmetry)
are sufficiently complicated that they must be addressed numerically,
the others are amenable to an analytic treatment.  \citet{furl05-rec}
made an important step in this direction by showing how recombinations
affect the basic model: they have little effect until the bubbles
reach their Str{\" o}mgren radius relatively late in reionization.
The main purpose of this paper is to examine how several other
assumptions affect the major features of the model, especially those
emphasised in \S \ref{charsize}, and to generalise it where possible.
This is important because numerical simulations do not yet have the
dynamic range to resolve the galaxies responsible for ionizing the
universe while simultaneously sampling a large enough volume to trace
the growth of \hii regions.  These relatively simple analytic tests of
the model offer the best hope of testing its applicability,
although complete simulations will eventually be necessary
to understand the entire process.  At the same time, we will take
advantage of the greatest strength of analytic models -- their
flexibility -- to examine how basic properties of the galaxy
population affect reionization and hence to isolate those properties
to which observations will be most sensitive.  We will find that the
FZH04 model is surprisingly robust: its key features remain valid over
a wide range in parameter space.  Analytic models will thus play an
important role in interpreting the evolving topology of neutral and
ionized gas throughout this epoch.

We consider several modifications to the basic formalism in \S
\ref{mods}.  We first allow a different halo mass function and a
mass-dependent ionizing efficiency.  We show that, although such
modifications can introduce measurable differences to the bubble
distribution, they do not affect the qualitative aspects of the model.
Then, in \S \ref{flucs}, we investigate whether stochastic
fluctuations in the galaxy population can significantly affect the
bubble size distribution.  In \S \ref{recomb} we show how these factors
affect the bubbles if recombinations are
included.  We give some examples of how our results affect observables
in \S \ref{obs}, and we conclude in \S \ref{disc}.

In our numerical calculations, we assume a cosmology with $\Omega_m=0.3$, $\Omega_\Lambda=0.7$, $\Omega_b=0.046$, $H=100 h \hunits$ (with $h=0.7$), $n=1$, and $\sigma_8=0.9$, consistent with the most recent measurements \citep{spergel03}.  We use the transfer function of \citet{eisenstein98} to compute the matter power spectrum.  Unless otherwise specified, we quote all distances in comoving units.

\section{The FZH04 Model}
\label{fzh}

The basic motivation of the model is to describe how \hii regions grow
and overlap around overdensities containing galaxies.  We begin by assuming
that a galaxy of mass $m_{\rm gal}$ can ionize a mass $\zeta m_{\rm
gal}$, where $\zeta$ is a constant that depends on (among other
things) the efficiency of ionizing photon production, the escape
fraction of these photons from the host galaxy, the star formation
efficiency, and the mean number of recombinations.  Each of these
quantities is highly uncertain (e.g., massive Population III stars can
dramatically increase the ionizing efficiency; \citealt{bromm01-vms}),
but at least to a rough approximation they can be collapsed into this
single efficiency factor.  The criterion for a region to be ionized by
the galaxies contained inside it is then $\fcoll > \zeta^{-1}$, where
$f_{\rm coll}$ is the fraction of mass bound in haloes above some
$\mmin$.  Unless otherwise specified, we will assume that $\mmin=m_4$,
corresponding to a virial temperature $T_{\rm vir}=10^4 \kel$, at
which point hydrogen line cooling becomes efficient.  In the extended
Press-Schechter model \citep{lacey93}, this places a condition on the
mean overdensity within a region of mass $m$, \bq \delta_m \ge
\delta_x(m,z) \equiv \delta_c(z) - \sqrt{2} K_\zeta [\sigma^2_{\rm
min} - \sigma^2(m)]^{1/2},
\label{eq:deltax}
\eq
where $K_\zeta \equiv {\rm erf}^{-1}(1 - \zeta^{-1})$, $\sigma^2(m)$ is the variance of density fluctuations on the scale $m$, $\sigma^2_{\rm min} \equiv \sigma^2(m_{\rm min})$, and $\delta_c(z)$ is the critical density for collapse.  The {\it global} ionized fraction is $\bxio=\zeta f_{\rm coll,g}$, where $f_{\rm coll,g}$ is the mean collapse fraction throughout the universe.

FZH04 showed how to construct the mass function of \hii regions from $\delta_x$ in an analogous way to the halo mass function \citep{bond91}.  The barrier in equation (\ref{eq:deltax}) is well approximated by a linear function in $\sigma^2$, $\delta_x \approx B(m,z) \equiv B_0 + B_1 \sigma^2(m)$. In that case the mass function of ionized bubbles (i.e., the comoving number density of \hii regions with masses in the range $m \pm \deriv m/2$) has an analytic solution \citep{sheth98,mcquinn05}:
\bqa
n_b(m) \, \deriv m & = & \sqrt{\frac{2}{\pi}} \ \frac{\bar{\rho}}{m^2} \ \left|
  \frac{\deriv \ln \sigma}{\deriv \ln m} \right| \ \frac{B_0(z)}{\sigma(m)} \nonumber \\
  & & \times \exp \left[ - \frac{B^2(m,z)}{2 \sigma^2(m)} \right] \, \deriv m,
\label{eq:nbm}
\eqa
where $\bar\rho$ is the mean density of the universe.  This should be compared to the analogous mass function for virialised dark matter haloes \citep{press74}:
\bqa
n_h(m) \, \deriv m & = & \sqrt{\frac{2}{\pi}} \ \frac{\bar{\rho}}{m^2} \ \left|
  \frac{\deriv \ln \sigma}{\deriv \ln m} \right| \ \frac{\delta_c(z)}{\sigma(m)} \nonumber \\ 
  & & \times \exp \left[-\frac{\delta_c^2(z)}{2 \sigma^2(m)} \right] \, \deriv m.
\label{eq:nhm}
\eqa
Note the close structural similarity between these two functions.  Note also that, while every dark matter particle must be part of a virialized halo, only a fraction $\bxio$ should be incorporated into bubbles.  The integral of equation (\ref{eq:nbm}) is slightly different from this expected value because of the linear barrier approximation (see eq.~11 in \citealt{mcquinn05}), but the error is typically $\la 1\%$.

Several key points of the FZH04 model deserve emphasis; for concrete
examples of these properties see, e.g., Figure~\ref{fig:dndr} below.
First, the bubbles are large: tens of comoving Mpc for $\bxio \ga
0.65$.  Furthermore, the bubbles attain a well-defined characteristic
size $R_c$ at any point during reionization.  The distribution of
bubble sizes also narrows considerably throughout reionization.
Finally, $n_b(m)$ varies only weakly with $\zeta$ for a fixed $\bxio$
(or equivalently it is nearly independent of the redshift of
reionization).  In the remainder of this section, we will examine the
physical origin of these properties.

\subsection{The characteristic bubble size}
\label{charsize}

The characteristic bubble size $R_c$ plays an analogous role to the ``break" mass $m_\star$ in the halo mass function.  This is simply the peak of $m^2 n_h(m)$, corresponding to the halo size that contains the most fractional mass.  If we assume that $| \deriv \ln \sigma/\deriv \ln m|=$constant [which would follow from a pure power-law matter power spectrum $P(k) \propto k^n$], we find
\bq
\sigma(m_\star) = \delta_c(z).
\label{eq:mstar}
\eq
In other words, the characteristic halo mass is simply the point at which the exponential cutoff sets in.  Physically, it is the scale at which a ``typical" 1-$\sigma$ density fluctuation will collapse on its own; this scale grows with time because $\delta_c(z)$ decreases.

We can define $R_c$ in exactly the same way.  Again assuming a power-law matter power spectrum, we have
\bq
\sigma^2(R_c) = \frac{-1 + (1 + 4 B_0^2 B_1^2)^{1/2}}{2 B_1^2}.
\label{eq:sigpeak}
\eq
The physical meaning is somewhat obscured in this case, but it has two simple limiting forms.  First, if $4 B_0^2 B_1^2 \ll 1$, $\sigma(R_c) \approx B_0$.  This occurs when $\bxio \sim 1$ ($B_0$ must be small if most of the universe is to lie inside \hii regions).  In this regime, the characteristic scale is simply the mass at which a typical 1-$\sigma$ trajectory crosses the barrier.  This is completely analogous to $m_\star$.  The final step, then, is to understand the meaning of $B_0$.  From equation (\ref{eq:deltax}), it corresponds to the self-ionization condition at $m=\infty$:
\bq
\zeta \fcoll(\delta=B_0,\sigma=0) = 1.
\label{eq:b0defn}
\eq
In other words, it is the overdensity required for the sources inside a given large region to ionize that region.  Thus, $R_c$ is simply the scale on which a typical 1-$\sigma$ density fluctuation contains enough galaxies to ionize itself.

In the opposite regime, when $4 B_0^2 B_1^2 \gg 1$ (early in reionization), we have $\sigma(R_c) \approx B_0/B_1$.  In this case, the simple analogy to $m_\star$ is less helpful, because $R_c$ is determined from the set of \emph{atypical} trajectories that actually cross the barrier (constituting a fraction $\bxio$ of all possible trajectories).  The increasing barrier selects a preferred scale from this set that depends on both $B_0$ and the slope of the barrier.

\begin{figure}
\begin{center}
\resizebox{8cm}{!}{\includegraphics{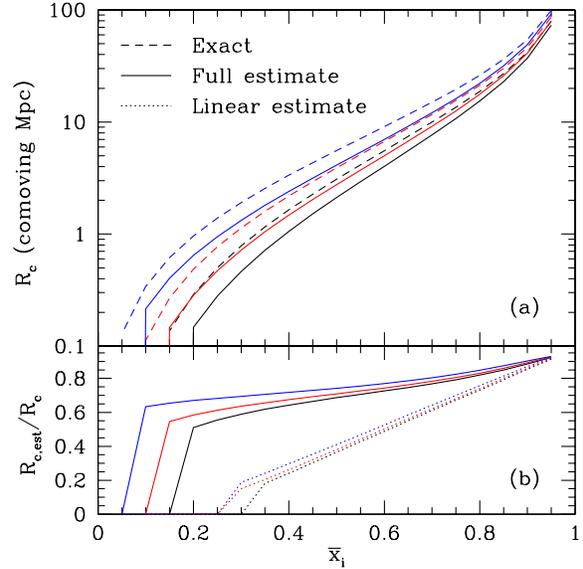}}\\%
\end{center}
%\plotone{rchar.eps}
\caption{\emph{(a)}: Characteristic bubble size.  Dashed curves show the exact $R_c$ for $z=14,\,10$, and $8$ from top to bottom.  Solid curves show the estimate of eq. (\ref{eq:sigpeak}).  \emph{(b)}:  Ratio of the estimated characteristic size to the true value.  The solid curves use eq. (\ref{eq:sigpeak}), while the dotted curves include only linear bias corrections (eq. \ref{eq:b0bias}).  }
\label{fig:rchar}
\end{figure}

Given $\bxio$ and $z$, we can invert equations (\ref{eq:b0defn}) and (\ref{eq:sigpeak}) to find $R_c$.  Figure~\ref{fig:rchar}\emph{a} compares the result (solid curves) to the exact $R_c$ (dashed curves) at $z=14,\,10,$ and $8$ (from top to bottom).  The solid curves in Figure~\ref{fig:rchar}\emph{b} show their ratio.  Our simple estimate matches the true result quite closely, to within a factor of two for $\bxio \ga 0.2$.  (The cutoff at small $\bxio$ occurs because the characteristic mass approaches $\zeta \mmin$, the smallest physically allowed \hii region.)  The estimate is not perfect because we have neglected the curvature of the cold dark matter (CDM) power spectrum; it worsens at smaller $\bxio$ because the trajectories sample a larger range of masses and also at lower redshifts because the galaxy distribution is more evolved.  

%%%%%%%%%%%FIGURE 1:  Characteristic Radius

We can go one step further in order to understand how $R_c$ depends on the ionizing source population.  To linear order we can write
\bq
\fcoll(\delta,\sigma=0) = \int \deriv m \, m \, n_h(m) \, [ 1 + \delta_z \, b_L(m)],
\label{eq:fcoll-lin}
\eq
where $\delta_z$ is the physical overdensity at redshift $z$ and $b_L(m)$ is the Lagrangian linear bias \citep{mo96}:
\bq
b_L(m) = \frac{\delta_c^2(z)/\sigma^2(m)-1}{\delta_c^0},
\label{eq:bm}
\eq
with $\delta_c^0=\delta_c(z=0)$.  Then equation (\ref{eq:b0defn}) becomes
\bq
B_0 \approx \frac{\bxio^{-1} - 1}{D(z) \, b_{\rm eff}},
\label{eq:b0bias}
\eq
where $D(z)$ is the linear growth factor [$D \propto (1+z)^{-1}$ at high redshifts] and 
\bq
b_{\rm eff} \equiv \frac{1}{\bar{\rho} \, f_{\rm coll,g}} \int \deriv m \, m \, b_L(m) \, n_h(m).
\label{eq:beff}
\eq  
The dotted curves in Figure~\ref{fig:rchar}\emph{b} show the results of this approximation (scaled to the exact solution).  The poor accuracy of the linear theory estimate may seem surprising given that $f_{\rm coll,g} \ll 1$.  One reason is that finite size effects suppress the galaxy population when $R_c \la 10 \Mpc$.  But a more important reason is cosmological:  the shape of the CDM power spectrum is such that $\sigma^2 \propto R^{-(3+n)}$ (approximately), with $n \rightarrow 1$ for $R \rightarrow \infty$ and $n \rightarrow -3$ for $R \rightarrow 0$.  The bubbles sample a wide range of effective slopes as reionization proceeds; we have $n \approx -2.3,\,-1.8$, and $-0.6$ at $R_c = 1,\,10$, and $100 \Mpc$.  Thus, early in reionization the radius is extremely sensitive to $\sigma$; in this regime, small errors in $B_0$ correspond to large errors in $R_c$.  At the later stages, however, $R_c$ becomes more robust to the approximations and the estimate is more accurate.

Although the $b_{\rm eff}$ approximation does not provide a fully satisfactory estimate for $R_c$, it is accurate to about a factor of two when $\bxio \ga 0.5$.  It also provides an important clue to the mechanism fixing the characteristic scale:  for a given $\bxio$, the crucial parameter is $D(z) b_{\rm eff}$.  More biased galaxy populations will drive reionization toward larger scales because they cluster more strongly.  The galaxies give a substantial ``boost" to each mass fluctuation and allow larger scales to cross the reionization barrier.  Thus, weighting massive galaxies more heavily will generically create larger bubbles.  Furthermore, equation (\ref{eq:b0bias}) explains why $n_b(m)$ is only weakly dependent on redshift for a fixed $\bxio$:  as noted by \citet{oh99}, the halo bias is such that the galaxy correlation length evolves fairly slowly from $z \sim 6$--$20$.  A similar factor $D(z) b_{\rm eff}$ determines $R_c$ here.

Finally, we note that the mere existence of $R_c$ does not explain why $n_b(m)$ has a small-scale cutoff (the halo mass function, for example, becomes a power law for $m \ll m_\star$).  In the excursion set formalism, the cutoff occurs because $\delta_x$ increases linearly with $\sigma^2$, so that random walks are not guaranteed to cross the barrier at any scale.  Physically, the barrier increases because the collapse fraction is suppressed as we move to smaller scales, because fewer modes are available and because of finite size effects.  

\subsection{The width of the bubble distribution}
\label{width}

The other characteristic we wish to explain is why $n_b(m)$ narrows as $\bxio \rightarrow 1$.  To understand this, we first consider $n_b(m)$ if $P(k) \propto k^n$; in this case $\bxio$ affects the logarithmic width of the distribution only through the barrier in the exponential factor (see eq. \ref{eq:nbm}).  In fact, the distribution of bubble sizes \emph{narrows} as $\bxio \rightarrow 0$ in such a cosmology:  when $\bxio$ is small, the dominant factor is $\exp[-B_0^2/(2\sigma^2)]$, and the distribution decays exponentially at small masses.  It actually evolves somewhat more slowly than this single factor indicates, because the other terms in the exponential compensate:  it even remains nearly invariant for $\bxio \ga 0.5$.  We find that, for a power law $P(k)$, the distribution decays exponentially with a characteristic non-dimensional length $\sigma^2(R_s)/\sigma^2(R_c) \equiv K \ga 3$, where $R_s$ is a small-scale cutoff, with $K$ increasing slowly as $\bxio$ increases.  This is the opposite of what we find in a CDM cosmology, where the distribution \emph{widens} at early times.

The fundamental reason for the narrowing at large $R_c$ must therefore
be the shape of the CDM power spectrum.  Again, we note that $\sigma^2
\propto R^{-(3+n)}$ with $n \approx -2.3,\,-1.8$, and $-0.6$ at $R_c =
1,\,10$, and $100 \Mpc$.  Thus, $R_s/R_c \approx K^{-1/(n+3)}$.  When the bubbles are large, $R_s/R_c \approx
K^{-0.4}$; the two scales must lie close to one another because the
variance changes rapidly with scale.  On the other hand, when $\bxio
\ll 1$ and $R_c \sim 1 \Mpc$, we find $R_s/R_c \approx K^{-1.4}$.
Thus, $R_s \ll R_c$, and the distribution is much broader at early
times.

Interestingly, this means that the underlying power spectrum has a
substantial effect on the growth of \hii regions during reionization.
Given the large number of astrophysical uncertainties in any model of
reionization, it seems unlikely that such behaviour can actually be
used as a probe of the power spectrum.  However, it does imply that
the existence of a well-defined characteristic scale is a generic
feature of \emph{any} reionization model based on the underlying
galaxy fluctuations in a CDM model.

\section{Generalisations of the FZH04 Formalism}
\label{mods}

We will now modify the FZH04 model in a number of useful ways.  We will see that the features described in \S \ref{fzh} are surprisingly robust to many of the simplifying assumptions. 

\subsection{A mass-dependent efficiency $\zeta(m)$}
\label{zetam}

One obvious simplification of FZH04 was to take the ionizing efficiency parameter $\zeta$ to be independent of galaxy mass.  In the local universe the star formation efficiency $f_\star \propto m^{2/3}$ in low-mass galaxies \citep{kauffmann03}; similar behavior could certainly occur for any of the parameters relevant to reionization, especially if they are driven by feedback.  Here we will examine a general class of models in which $\zeta=\zeta(m)$; we will take $\zeta \propto m^\alpha$ in our numerical calculations.  To do so we simply replace the condition $\zeta \fcoll=1$ with
\bq
\int \deriv m \, \zeta(m) \, m \, n_h(m|\delta,M) = 1,
\label{eq:zetam}
\eq
where $n_h(m|\delta,M)$ is the halo mass function within a large region of mass $M$ (or equivalently radius $R$) and smoothed overdensity $\delta$.  According to the extended Press-Schechter model, this is \citep{lacey93}
\bqa
n_h(m|\delta,M) & = & \sqrt{\frac{2}{\pi}} \ \frac{\bar{\rho}}{m^2} \ \left|
  \frac{\deriv \ln \sigma}{\deriv \ln m} \right| \ \frac{\sigma^2[\delta_c(z)-\delta]}{[\sigma^2(m)-\sigma^2(M)]^{3/2}} \nonumber \\
& & \times  \exp \left\{-\frac{[\delta_c(z)-\delta]^2}{2 [\sigma^2(m)-\sigma^2(M)]} \right\}.
\label{eq:nhcond}
\eqa
The parallel with the original condition is more obvious if we define $\bar{\zeta}=\bar{\zeta}(\delta,M)$ to be the mass-weighted ionizing efficiency within the specified region; then equation (\ref{eq:zetam}) becomes $\bar{\zeta} \fcoll =1$.  This condition allows us to find (numerically) the minimum $\delta_x$ for a region to self-ionize, just as in equation (\ref{eq:deltax}), and hence the appropriate $n_b(m)$ (note that the barriers remain nearly linear in $\sigma^2$).

\begin{figure}
\begin{center}
\resizebox{8cm}{!}{\includegraphics{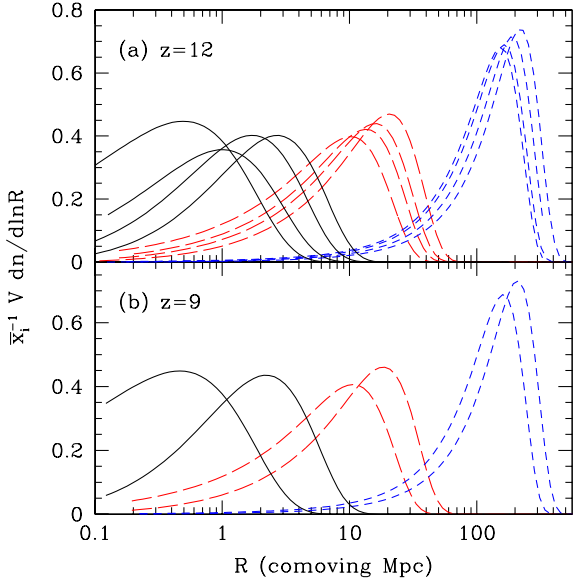}}\\%
\end{center}
%\plotone{dndr_fstar.eps}
\caption{ Bubble size distributions for $\zeta \propto m^\alpha$.  The solid, long-dashed, and short-dashed sets have $\bxio=0.16,\,0.65$, and $0.975$.  \emph{(a)}: $z=12$.  Within each set, the curves assume $\alpha = 0,\,1/3,\,2/3$, and $1$, from left to right. \emph{(b)}: $z=9$.  We show $\alpha=0$ and $2/3$ in this case.} 
\label{fig:dndr}
\end{figure}

Figure~\ref{fig:dndr} shows the resulting size distributions.  The solid, long-dashed, and short-dashed curves have $\bxio=0.16,\,0.65$, and $0.975$.  In panel \emph{(a)}, we show the distributions at $z=12$.  Within each set, the curves have $\alpha=0,\,1/3,\,2/3$, and $1$, from left to right.  The most important result is that $R_c$ increases with $\alpha$.  This is to be expected from equation (\ref{eq:b0bias}):  massive galaxies are more highly clustered, so they drive reionization to larger scales.  The degree of amplification decreases as $\bxio \rightarrow 1$ because of the shape of the power spectrum, just as described in \S \ref{width}.  At early times, there is an additional contribution because the $\zeta(m)$ excursion set barrier $\delta_x$ is steeper, which moves the peak to smaller $\sigma$ or large scales [see the discussion following equation (\ref{eq:sigpeak})].  The steepening occurs because more photons come from massive galaxies, whose abundances are more efficiently suppressed on smaller scales.  Note, however, that $R_c$ never increases by more than a factor of a few:  this is because the barriers tend to intersect fairly near the characteristic scale.  Panel \emph{(b)} shows the distributions at $z=9$.\footnote{Note that we take $\mmin=3.3 m_4$ here in order to force $\zeta>1$ in all star-forming haloes.}  Clearly the differences are comparable to the higher redshift case.

%%%%%%%%%%%%%FIGURE 2:  SF Mass Functions

\subsection{The halo mass function}
\label{nh}

While the Press-Schechter mass function is elegant and provides a reasonable description of halo populations in numerical simulations, it is possible to improve the agreement at lower redshifts.  The \citet{sheth99} mass function provides a better fit to the simulation results \citep{jenkins01}:
\bqa
n_{\rm ST}(m) \, \deriv m & = & A \sqrt{\frac{2 a}{\pi}} \ \frac{\bar{\rho}}{m^2} \ \left|
  \frac{\deriv \ln \sigma}{\deriv \ln m} \right| \ \left\{ 1 + \left[ \frac{a \delta_c^2(z)}{\sigma^2} \right]^{-p} \right\} \nonumber \\
  & & \times \frac{\delta_c(z)}{\sigma(m)} \ \exp \left[-\frac{\delta_c^2(z)}{2 \sigma^2(m)} \right] \deriv m,
\label{eq:nhst}
\eqa
where $A=0.322$, $a=0.707$, and $p=0.3$.  Although originally motivated through fits to numerical simulations, this form also has some physical justification.  \citet{sheth01}  showed that the dynamics of ellipsoidal collapse yield an excursion set barrier whose mass function is similar to equation (\ref{eq:nhst}).  At high redshifts, the mass function is less well-known, but it seems to lie somewhere in between these two versions \citep{jang01,reed03}.  We will now generalise FZH04 to include the Sheth-Tormen mass function.

The first effect of the new mass function is to change $f_{\rm coll,g}$:  it is larger than the Press-Schechter value at these redshifts because the Sheth-Tormen function yields more high-mass objects on the exponential tail of the halo distribution.  This alone does not, however, imply that $n_b(m)$ will change:  it simply means that the two mass functions require different mean efficiencies $\zeta$.  The size distributions will change only if the local collapse fraction behaves differently; i.e. whether the solution to $\zeta \fcollst(\delta,\sigma)=1$ has different scale of density dependence.  Thus, we need to compute the Sheth-Tormen mass function in an arbitrary region of overdensity $\delta$.  Unfortunately, the form in equation (\ref{eq:nhst}) cannot easily be extended to finite regions; in contrast to flat and linear barriers, the relevant excursion set barrier changes shape after a shift of the origin.  \citet{sheth02} found that one could approximate the conditional mass function $n_{\rm ST}(m|\delta,M)$ through a Taylor series expansion of the barrier after a translation of the origin.  We will use this approximate form; it should suffice to illustrate the sensitivity to uncertainties in the mass function.\footnote{We do note one subtlety with this method:  the Taylor-series fit to the mass function provided by \citet{sheth02} does not agree exactly with equation (\ref{eq:nhst}) and gives a different mean collapse fraction.  We therefore renormalise all of the conditional $\fcollst$ by the ratio of the mean fractions.}  We note that \citet{barkana04-fluc} took the Sheth-Tormen abundances to vary with $\delta$ in precisely the same way as the Press-Schechter function does; if this is a reasonable approximation (as it appears to be from their comparison to simulations), we would expect $n_b(m)$ to be independent of the halo mass function.

\begin{figure}
\begin{center}
\resizebox{8cm}{!}{\includegraphics{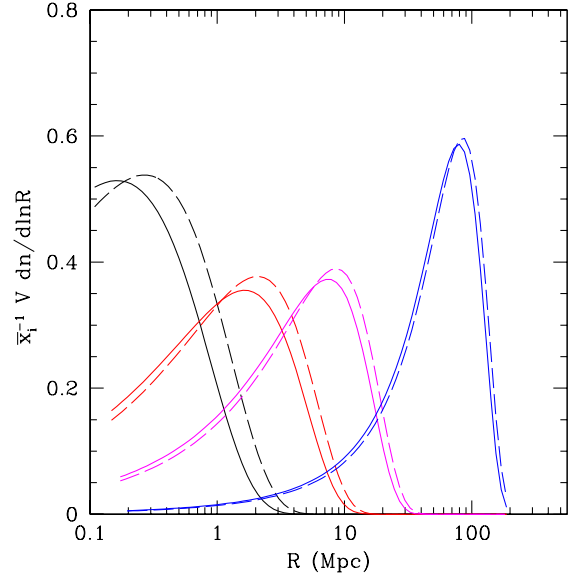}}\\%
\end{center}
%\plotone{dndr_st.eps}
\caption{ Bubble size distributions for different halo mass functions at $z=8$.  From left to right, the sets of curves take $\bxio=0.16,\,0.4,\,0.65$, and $0.95$; within each pair, the solid and dashed curves use the Press-Schechter and Sheth-Tormen halo mass functions, respectively.}
\label{fig:dndrst}
\end{figure}

Figure~\ref{fig:dndrst} shows the resulting size distributions at $z=8$.  Clearly the halo mass function has only a subtle effect on $n_b(m)$.  The slight increase in $R_c$ occurs because the Sheth-Tormen mass function contains more massive, biased haloes.  The corresponding excursion set barriers are nearly coincident; only when $\bxio$ is small do they differ appreciably.  At higher redshifts, the differences are even smaller than shown in this figure (at these times the mass functions are so steep that slight differences in its shape cannot have any significant effect).  Thus it does not appear that uncertainties in the halo mass function will significantly modify the bubble distribution.

%%%%%%%%%%%FIGURE 3:  nh Mass Functions

For completeness, we also note that \citet{mcquinn05} discussed some
related issues.  They found that the mass function -- and even the
filtering scheme -- can have a significant impact on the duration of
reionization because it affects $f_{\rm coll,g}$ (and its time derivative).
Here we have shown that $n_b(m)$ is nearly independent of the halo
mass function at a fixed $\bxio$.

\subsection{Stochastic fluctuations in the galaxy distribution}
\label{flucs}

To this point, we have assumed that the mapping between the mean
matter density and the galaxy density is exact, so that the ionized
bubbles trace the large-scale structure in a deterministic fashion.
However, in reality the galaxy distribution will fluctuate.  This may
be important, because the FZH04 model shows that -- particularly
toward the end of reionization -- slight overdensities in the galaxy
distribution can strongly affect the bubble sizes.  This is a
manifestation of the ``phase transition" nature of reionization.
\citet{sheth-lemson99} and \citet{casas02} studied this ``stochastic
bias" in some detail.  They found that the variance of galaxy counts
in simulations of the $z=0$ universe is nearly Poissonian in underdense
regions, somewhat smaller in regions near the mean density, and
somewhat larger in overdense regions.  They explained the variation
through finite size effects and clustering.  We will consequently
assume that the halo number counts at each mass obey Poisson
statistics.  Most of the ionized regions have only modest
overdensities, so comparison to the lower-redshift simulations
suggests that we may overestimate the variance by a factor of order
unity; \citet{casas02} found the discrepancy relative to Poisson to be
no more than a factor of two.

How will such fluctuations affect $n_b(m)$?  Physically, the FZH04 condition $\zeta \fcoll=1$ compares the number of ionizing photons to the number of hydrogen atoms within a specified region.  Thus the relevant figure of merit for estimating the accuracy of our prescription is the variance in the number of ionizing photons produced inside a bubble, $\sigma_N^2$.  Each galaxy ionizes a number of atoms equal to $N_{i,g} = \zeta m_g/m_H$, where $m_g$ is its mass and $m_H$ is the proton mass.  If all galaxies had the same mass, the variance would be $\sigma_N^2 = (\zeta m_g/m_H)^2 \, \bar{n}_h V$ (assuming Poisson statistics), where $\bar{n}_h$ is the mean galaxy number density and $V$ is the volume of a region with mass $M$.  Allowing for a range of galaxy masses, we have
\bq
\sigma_N^2(M) = V \ \int \deriv m \ \zeta^2(m) \ \frac{m^2}{m_H^2} \ n_h(m|\delta_x,M).
\label{eq:vareps}
\eq
Fluctuations in the galaxy distribution correspond to changes in the number of ionizing photons per baryon.  We can therefore quantify their effects in terms of fractional fluctuations in the local ionizing efficiency $\zeta$:
\bq
\delta_\zeta \equiv \sigma_N/\langle N_i \rangle,
\label{eq:deltaz}
\eq
where $\langle N_i \rangle = M/m_H$ is the expectation value of the total number of ionizing photons.

\begin{figure}
\begin{center}
\resizebox{8cm}{!}{\includegraphics{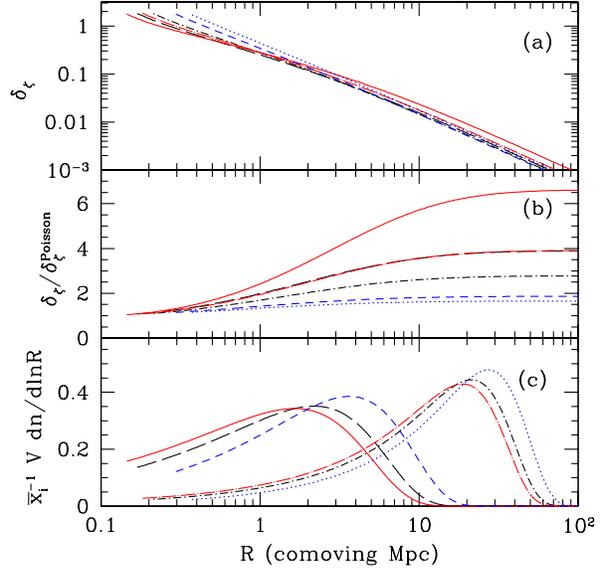}}\\%
\end{center}
%\plotone{fluc_zevol.eps}
\caption{\emph{(a)}:  Relative fluctuations in the number of photons per baryon $\delta_\zeta$.  \emph{(b)}:  Ratio of fluctuations in $\zeta$ to those expected from number counts alone.  \emph{(c)}:  Bubble size distributions according to the FZH04 model (neglecting stochastic fluctuations; these are included only for reference).  Each curve takes $\zeta=$constant.  The solid, long-dashed, and short-dashed curves assume $\bxio=0.4$ at $z=8,\,10$, and $15$, respectively.  The long dot-dashed, short dot-dashed, and dotted curves show the same redshifts but for $\bxio=0.8$. }
\label{fig:fluczevol}
\end{figure}

Figure~\ref{fig:fluczevol} illustrates the importance of these fluctuations.  For reference, panel \emph{(c)} shows $n_b(m)$ in the standard FZH04 model for several different $\bxio$ and redshifts.  Panel \emph{(a)} shows $\delta_\zeta$ as a function of scale.  For this fiducial model, the fluctuations are of order unity for $R \la 0.3 \Mpc$ and fall below $10\%$ for $R \ga 3 \Mpc$.  Because the bubbles pass this threshold fairly early in reionization, they should not significantly alter $n_b(m)$ unless $\bxio$ is relatively small.

%%%%%%%%%%: FIGURE 4: Stochastic fluctuations

Interestingly, the fluctuations are nearly independent of both redshift and $\bxio$.  To understand this behaviour, it is useful to rewrite equation (\ref{eq:deltaz}) as
\bq
\delta_\zeta = \frac{1}{\sqrt{\bar{n}_h V}} \ \frac{\langle m^2 \rangle^{1/2}}{\langle m \rangle},
\label{eq:dzetm}
\eq
where we have assumed $\zeta=$ constant and the averages are over the conditional mass function.  The first factor is the variance $\delta_\zeta^{\rm Poisson}=N_{\rm gal}^{-1/2}$ expected from pure Poisson fluctuations; the second describes the distribution of galaxies.  Note that both factors implicitly depend on the bubble mass $M$.  Let us first focus on $\delta_\zeta^{\rm Poisson}$ and assume that all galaxies have the same mass.  Then (for a given bubble size) $N_{\rm gal} \propto 1/\zeta$ and $\delta_\zeta^{\rm Poisson} \propto \zeta^{1/2}$; clearly in this simple case the fluctuations would evolve with redshift (and $\bxio$).

However, in Figure~\ref{fig:fluczevol}\emph{b} we show that, because of the mass function, $\delta_\zeta^{\rm Poisson}$ does not necessarily dominate unless the bubbles are small (so that $N_{\rm gal} \la 1$).  The range of galaxy masses increases with cosmic time, increasing $\langle m^2 \rangle$ and hence $\delta_\zeta$ relative to the Poisson expectation.  Evidently, evolution in the shape of the mass function (specifically $\mmin/m_\star$) compensates for the increased number of galaxies at lower redshifts, making $\delta_\zeta$ more or less constant with redshift.  

Equation (\ref{eq:deltaz}) also helps illuminate why $\delta_\zeta$ changes only slowly with $\bxio$ at a fixed redshift.  If the bubbles always took the mean density, their interior mass functions would be independent of $\zeta$ and both factors in equation (\ref{eq:deltaz}) would be independent of $\bxio$.  However, the bubbles actually sit at density $\delta_x$, which is a function of $\bxio$; as a result the shape of the mass function in ionized regions changes slightly because bubbles are more atypical regions when $\bxio$ is small.  This also explains why $\delta_\zeta/\delta_\zeta^{\rm Poisson}$ is constant in large bubbles:  all have nearly the same $\delta_x \approx B_0$ and $\sigma \ll 1$, so $n_h(m|\delta_x,M)$ approaches a constant shape.  In this regime, $\delta_\zeta \propto N_{\rm gal}^{-1/2} \propto R^{-3/2}$. 

The fluctuations approach $\delta_\zeta^{\rm Poisson}$ in regions where the mass function is steep; in that case most galaxies have $m \approx \mmin$.  We see this in the $z=15$ curves in Figure~\ref{fig:fluczevol}\emph{b}.  It also happens if we increase $\mmin$, pushing galaxies farther off on the exponential tail of the mass function.  However, in that case $\delta_\zeta$ increases because $N_{\rm gal}$ falls by an even larger factor. 

\begin{figure}
\begin{center}
\resizebox{8cm}{!}{\includegraphics{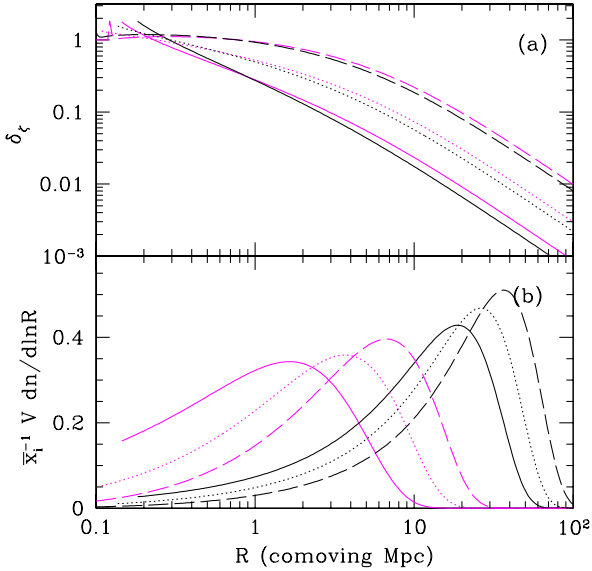}}\\%
\end{center}
%\plotone{fluc_z8_sf.eps}
\caption{\emph{(a)}:  ionizing efficiency fluctuations $\delta_\zeta$.  \emph{(b)}: Bubble size distributions at $z=8$ (ignoring stochastic fluctuations).  The solid, dotted, and dashed curves assume $\zeta \propto m^\alpha$, with $\alpha=0,\,1/3$, and $2/3$, respectively.  The left and right sets take $\bxio=0.4$ and $0.8$.  }
\label{fig:flucz8sf}
\end{figure}

Another way to increase $\delta_\zeta$ is to take $\zeta \propto m^\alpha$, with $\alpha > 0$.  
Figure~\ref{fig:flucz8sf} shows the fluctuations if $\alpha = 0,\,1/3$, and $2/3$ (solid, dotted, and dashed curves, respectively) at $z=8$.  Panel \emph{(a)} shows that the fluctuations can be quite strong when the efficiency is a steep function of galaxy mass, exceeding $10\%$ for $R \la 20 \Mpc$.  In this case the second factor in equation (\ref{eq:deltaz}) becomes $\langle \zeta^2 m^2 \rangle^{1/2}/\langle \zeta m \rangle$, which is large ($\sim 20$ and $60$ for $\alpha = 1/3$ and $2/3$) because it exaggerates the importance of massive galaxies.  Evidently, so few sources contribute  that our deterministic approximation may break down until quite late in the process.   

%%%%%%%%%%%%%FIGURE 5: Fluctuations and zeta(m)

How will these fluctuations affect $n_b(m)$?  Clearly our analytic model breaks down if $\delta_\zeta \sim 1$, and numerical simulations will then be necessary to study the process in detail.  Nevertheless, we can guess the qualitative effects.  Consider a region that would nominally form a bubble of volume $V_0$ (corresponding to a density threshold $\delta_x$).  First, suppose that it has fewer galaxies than expected, with an efficiency $\zeta_- < \zeta$.  In the excursion set formalism, this implies an effective barrier height $\delta_x' > \delta_x$.  We could therefore estimate the local bubble distribution by considering random walks beginning at $(\delta_x,V_0)$ and computing where they cross the $\delta_x'$ barrier.  This is essentially equivalent to searching for the progenitor distribution at a slightly earlier redshift.  \citet{furl05-rec} showed that most of the volume is contained in a single progenitor slightly smaller than $V_0$.  Thus, negative fluctuations will smear the volume distribution by a factor $\delta_V \approx \delta_\zeta$, where $\delta_V = (V-V_0)/V_0$.

Positive fluctuations are slightly more subtle because bubbles can merge.  Ignoring clustering, each \hii region is surrounded by neutral gas with volume $\approx V_0/\bxio$.  Early in reionization, these thick walls prevent the bubbles from merging and $\delta_V \approx \delta_\zeta$.  On the other hand, if $\delta_\zeta > (1/\bxio-1)$, the bubble will grow enough to touch its neighbor.  For larger fluctuations, bubbles will grow both by ionizing new gas and by consuming their neighbors; the volume fluctuation will then be $\delta_V \approx \delta_\zeta/(1-\bxio)$.  Because the extra sources need only ionize the thin walls between bubbles, a modest boost in the number of galaxies can greatly increase the bubble's size.  Thus, near the end of reionization, small fluctuations can broaden the size distribution by significantly more than $\delta_\zeta$.  Clustering will also increase the importance of merging.  However, during this epoch the bubble sizes are most likely limited by recombinations.  We now turn to that regime.

\section{Bubbles and Recombinations}
\label{recomb}

FZH04 included recombinations in only the crudest possible way, incorporating the mean number of recombinations per hydrogen atom into $\zeta$.  This assumption of a spatially uniform recombination rate is a gross oversimplification \citep[hereafter MHR00]{miralda00}:  the IGM is itself inhomogeneous, so the recombination time varies across voids, filaments, and collapsed objects.  MHR00 argued that (on a local level) reionization therefore proceeds from low to high densities.  

\citet{furl05-rec} showed how to reconcile this picture with the FZH04 model in which reionization begins around large-scale overdensities
\citep{ciardi03-sim, sokasian03,sokasian04}.  For a bubble to grow, the (local) mean free path of an ionizing photon must exceed its radius.  The mean free path is determined by the spatial extent and number density of dense, neutral blobs (where the recombination time is short).  Thus, as a bubble grows, its radiation background must also increase to ionize each of these blobs more deeply.  As a consequence, the recombination rate inside the bubble also increases; \hii regions cease growing when this recombination rate exceeds the ionizing emissivity (see also MHR00).  In other words, an ionized region must fulfill a second condition: 
\bq
\epsilon_{\rm bub} \equiv \zeta \frac{\deriv \fcoll(\delta,R)}{\deriv t} > A_{\rm bub}(R,\delta)
\label{eq:recomb-rate}
\eq
where $\epsilon_{\rm bub}$ is the normalised emissivity and $A_{\rm bub}(R,\delta)$ is the recombination rate within the bubble (which can be evaluated for a given IGM model).  The latter depends on the mean overdensity of the bubble through the average recombination rate and increases with the bubble radius.  This is analogous to the FZH04 condition $\zeta \fcoll>1$ except that it compares the instantaneous, rather than cumulative, number of photons.  \citet{furl05-rec} showed that, for reasonable models of the IGM density field, equation (\ref{eq:recomb-rate}) halts bubble growth at a well-defined radius $R_{\rm max}$; any bubbles nominally larger than this size fragment into regions with $R \approx R_{\rm max}$.  (Note, however, that this saturation radius describes the mean free path of ionizing photons, not the extent of contiguous ionized gas; obviously as $\bxio \rightarrow 1$ ionized gas does fill the universe.)  Unfortunately, calculating $R_{\rm max}$ requires knowledge of the IGM density field during reionization, which is currently an open question.  MHR00 present a numerical fit to the IGM density in cosmological simulations at $z=2$--$4$ that can be extrapolated to higher redshifts; for that model, $R_{\rm max} \ga 20 \Mpc$ (and much larger when $z>10$).  Thus, it only significantly affects $n_b(m)$ when $\bxio \ga 0.8$.  However, their model explicitly smooths the IGM density field on the Jeans scale of ionized gas, which may not be appropriate during reionization.  If smaller-scale structures such as minihaloes are common before reionization, recombinations may become important much earlier (e.g., \citealt{haiman01-mh,shapiro04,iliev05}).  For concreteness we will use the MHR00 IGM model in these examples.

\subsection{Stochastic fluctuations and recombinations}
\label{fluc-recomb}

The FZH04 photon counting argument uses the cumulative number of ionizing photons, so the relevant fluctuating quantity is the total number of galaxies in a region.  The recombination limit, on the other hand, depends on the number of \emph{active} sources at any time.  Even if the total number of galaxies is fixed, each could shine for only a fraction $p$ of the Hubble time (we will assume that $p$ is independent of mass for simplicity).  The variance in the luminosity of each galaxy is then $(1-p)/p$ times its mean square value; fluctuations in the emissivity are
\bq
\delta_\epsilon = \sqrt{\frac{1-p}{p}} \, \delta_\zeta.
\label{eq:deltaemiss}
\eq
In the limit of a small duty cycle, the emissivity fluctuations approach the number count fluctuations divided by $p^{1/2}$, which accounts for the reduced number of active sources.  The factor of $(1-p)$ in the numerator forces the fluctuations to vanish as the duty cycle approaches unity, because in that case all galaxies are active.\footnote{We have ignored fluctuations in the \emph{total} number of galaxies here, because we wish to consider emissivity fluctuations in a given bubble whose size is determined by the true number of galaxies, not its expectation value.  Including both sources requires the replacement $(1-p)/p \rightarrow 1/p$.}

%%%%%%%%%FIGURE 6: Fluctuations and recombinations
\begin{figure}
\begin{center}
\resizebox{8cm}{!}{\includegraphics{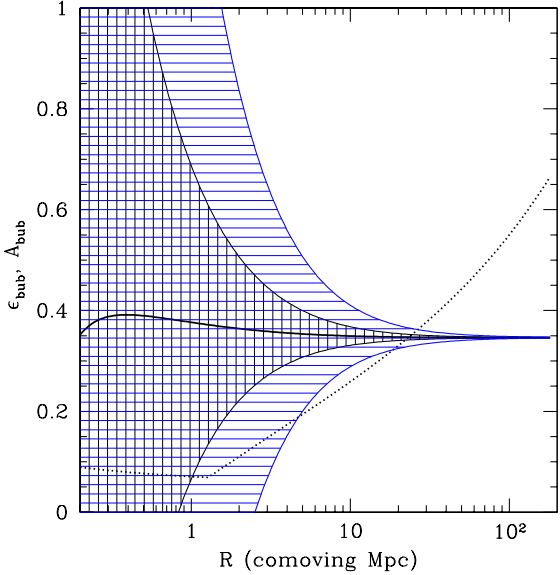}}\\%
\end{center}
%\plotone{rec_z8.eps}
\caption{Comparison of the emissivity inside a bubble (solid curve) to its recombination rate (dotted curve) as a function of bubble size at $z=8$ (assuming $\bxio=0.8$).  The shaded envelopes show the range of $\delta_\epsilon$.  The horizontally and vertically shaded regions take $p=0.01$ and $0.1$, respectively.}
\label{fig:recz8}
\end{figure}

Fluctuations in $\epsilon_{\rm bub}$ will affect the distribution of ionized and neutral gas within a bubble as well as the rate at which it grows.  If $\delta_\epsilon > 0$, it can ionize dense blobs more deeply, but that makes little difference because most photons can already reach the edge anyway.  If $\delta_\epsilon < 0$, the dense blobs will begin to recombine and more photons will be trapped inside.  However, the effects will not be serious unless $\epsilon_{\rm bub}$ falls below $A_{\rm bub}$; even in that case, the bubble only shrinks if the recombination time is shorter than the timescale of the fluctuations (i.e., in dense regions along the edge).  Fluctuations would be most interesting near $R_{\rm max}$, because they would then introduce an intrinsic scatter to the saturation radius, which is otherwise quite well-defined \citep{furl05-rec}.

Figure~\ref{fig:recz8} illustrates these effects.  The dotted curve shows $A_{\rm bub}$ (in units of recombinations per baryon per unit redshift) as a function of bubble scale at $z=8$ assuming $\bxio=0.8$.\footnote{The turnover at $R<1 \Mpc$ occurs because we force each bubble to be ionized up to its mean density, even if the formal limit from the mean free path is smaller.}  The thick curve show the emissivity $\epsilon_{\rm bub}$ (in units of ionizing photons per baryon per unit redshift).  Their intersection gives $R_{\rm max}$.  The shaded envelopes delineate $\epsilon_{\rm bub}  (1\pm \delta_\epsilon)$, illustrating the range of emissivity fluctuations.  The horizontal and vertical hatched regions take $p=0.01$ and $0.1$, respectively (note that the age of the universe is $t_H \approx 6.3 \times 10^8 \yr$ at $z=8$).  If the duty cycle is set by the dynamical time of galaxies (e.g., because active phases follow mergers), $p \approx (\bar{\rho}/\rho_{\rm vir})^{1/2} \approx 0.075$ and fluctuations are only important in small bubbles.  On the other hand, if $p \la 0.01$ (as may be appropriate for quasars), fluctuations can be significant even in relatively large bubbles.  We emphasise again, however, that the bubbles only shrink if \emph{both} $\epsilon_{\rm bub} < A_{\rm bub}$ and the recombination time is short compared to $p t_H$, which only happens for the mean density IGM when $z \ga 10$, even if $p \approx 1$.  Also, the fluctuations remain relatively small near $R_{\rm max}$ because of the enormous number of galaxies in these large bubbles.  Thus, emissivity fluctuations should not introduce a large spread in the saturation radius, and many of the conclusions of \citet{furl05-rec}, which relied on the existence of a sharply-defined characteristic radius, should remain valid.

\subsection{Recombinations and the $\zeta(m)$ barrier}
\label{mod-recomb}

%%%%%%%%%%FIGURE 7: Recombinations and massive galaxies
\begin{figure}
\begin{center}
\resizebox{8cm}{!}{\includegraphics{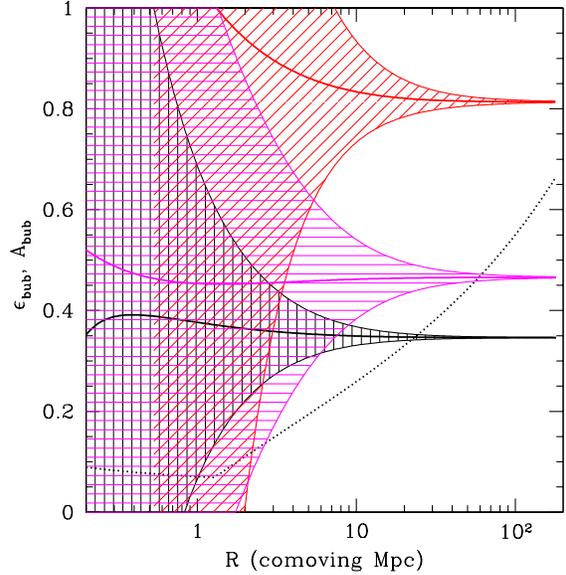}}\\%
\end{center}
%\plotone{rec_param.eps}
\caption{Comparison of the emissivity (solid curves) inside a bubble to the recombination rate (dotted curve) as a function of bubble size at $z=8$ (assuming $\bxio=0.8$).  The shaded envelopes show the range of $\delta_\epsilon$.  The bottom curve with vertical shading takes $\zeta=$constant and $\mmin=m_4$.  The middle curve with horizontal shading takes $\zeta \propto m^{1/3}$ and $\mmin=m_4$.  The uppermost curve with diagonal shading takes $\mmin=10m_4$ and $\zeta=$constant.  All the shaded regions assume $p=0.1$.}
\label{fig:recparam}
\end{figure}

We now consider recombinations in the presence of a mass-dependent ionizing efficiency.  This is easy to incorporate into \citet{furl05-rec}:  we simply replace the left-hand side of equation (\ref{eq:recomb-rate}) with $\deriv (\bar{\zeta} \fcoll)/ \deriv t$.  As mentioned in \S \ref{zetam}, weighting massive galaxies more heavily will increase the instantaneous emissivity at a fixed $\bxio$ because massive galaxies evolve more rapidly.  Thus saturation from recombinations will thus occur at larger scales.  Figure~\ref{fig:recparam} illustrates this point.  The bottom solid curve (surrounded by the vertical hatched envelope) takes $\zeta=$constant, while the middle curve (surrounded by the horizontal hatching) takes $\zeta \propto m^{1/3}$.

The ratio between the solid curves characterises the different rates at which the underlying galaxy populations evolve:  the relevant haloes grow $\sim 25\%$ faster for $\zeta \propto m^{1/3}$.  Thus the bubbles grow significantly larger when $\zeta \propto m^{1/3}$ than when it is a constant parameter.  For this example, $R_{\rm max} \approx 23,\,63$, and $180 \Mpc$ for $\alpha = 0,\,1/3$, and $2/3$.  Of course, $R_{\rm max}$ also depends sensitively on the IGM density (we have used the MHR00 model here), and these quantitative predictions are thus not reliable (causality will also limit the bubbles; \citealt{wyithe04-var}).  However, we have shown that whatever the underlying density structure, $R_{\rm max}$ increases significantly if massive galaxies drive reionization.  The uppermost curve (surrounded by the diagonally-hashed envelope), which takes  $\mmin=10 m_4$ and $\zeta=$constant, reinforces this point.  Increasing the threshold mass by this factor makes the emissivity so large that $R_{\rm max} \ga 300 \Mpc$ (at least in the MHR00 model).

Equation (\ref{eq:deltaemiss}) shows that $\delta_\epsilon \propto \delta_\zeta$, so the same factors that increase the number count variance will also increase that for the emissivity.  The envelopes in Figure~\ref{fig:recparam} illustrate this:  the relative fluctuations are largest for $\zeta \propto m^{1/3}$.  They are nearly Poisson for $\mmin = 10 m_4$, because most galaxies are close to the threshold and $\langle m^2 \rangle^{1/2} \approx \mmin$.  Note as well that by increasing $\epsilon_{\rm bub}$, scenarios relying on massive galaxies usually render emissivity fluctuations less important simply because $\epsilon_{\rm bub} \gg A_{\rm bub}$ on average.

\section{Observable Consequences}
\label{obs}

The issues we have considered in this paper all affect the bubble size distribution during reionization.  This is obviously one of the fundamental properties of any reionization scenario, and a number of
ways have been proposed to study it.  In this section we will briefly consider two of these techniques.  For concreteness, we will contrast the original FZH04 model and a scenario with $\zeta \propto m^{2/3}$.

\subsection{21 cm measurements}
\label{21cm}

Several low-frequency radio arrays designed to detect high redshift ($z \sim 6$--$20$) fluctuations in the 21 cm emission from neutral hydrogen are presently in the planning stages and/or under construction (e.g. PAST, LOFAR, MWA, and SKA).\footnote{For more information, see \citet{pen04}, http://www.lofar.org/, http://web.haystack.mit.edu/arrays/MWA/, and http://www.skatelescope.org/, respectively.}  These telescopes should answer many open questions about the reionization epoch.  They will measure fluctuations in $T_b$, the difference between the observed 21 cm brightness temperature and the CMB temperature $T_{\gamma}$ \citep{field59-obs}:
\bqa
T_b & \approx & 26 \, \left( \frac{T_s - T_\gamma}{T_s}\right)  ~ \left( \frac{\Omega_b h^2}{0.022} \right)~\left[\left(\frac{0.15}{\Omega_m h^2} \right)\, \left( \frac{1+z}{10}\right) \right]^{1/2} \nonumber \\
& & \times (1 + \delta)\, x_{\rm HI} \, ~~{\rm mK},
\label{eq:Tb2}
\eqa
where $T_s$ is the spin temperature and $x_{\rm HI}$ is the neutral fraction.  Equation (\ref{eq:Tb2}) ignores peculiar velocities, which tend to enhance the signal for modes along the line-of-sight
\citep{bharadwaj04-vel, barkana05-vel}.  Modes perpendicular to the line-of-sight are, for the most part, unaltered by the velocity field.  \citet{mcquinn05b} show that when the \hii regions are large, these redshift-space distortions are small on the large scales accessible to upcoming experiments.  We are interested in the 21 cm signal when the bubbles dominate, so we will ignore peculiar velocities.

It is likely that X-rays from the first stars, shocks and black holes heated the gas to temperatures well above $T_\gamma$ early in reionization \citep{oh01, venkatesan01, chen04}.  At the same time, Lyman-continuum photons from the first stars coupled the spin temperature $T_s$ to the gas temperature through the Wouthuysen-Field effect \citep{wouthuysen52, field58-wf}.  While this process was probably slower than the gas heating, \citet{ciardi03-21cm} argued that it becomes effective even when $\bxio \la 1\%$.  Therefore, we will take the limit $T_s \gg T_{\gamma}$, such that the 21 cm brightness temperature at the location ${\bf x}$ is given by
\begin{equation}
\Delta \, T_b({\bf x}) = \tilde{T}_b ~\left(1 -
\bar{x}_i \,[1 + \delta_x({\bf x})]\right) \left(1 + \delta({\bf
x}) \right) - \tilde{T}_b, \label{dTeqn}
\end{equation}
where $\tilde{T}_b \equiv \bar{T}_b/\bxh$ is the normalised mean brightness temperature, $\bxh \equiv 1 - \bxio$ is the global neutral fraction, $\delta_x$ is the overdensity in the ionized fraction, and $\delta$ is the matter overdensity (the baryons trace the dark matter on the relevant scales).

The imaging capabilities of upcoming observatories will be limited,
but statistical observations may still be quite powerful
\citep{zald04, bowman05}.  The simplest such statistic is the 3-D
power spectrum (FZH04).  Including terms up to second order in the
perturbations $\delta$, the brightness temperature power spectrum is
\begin{eqnarray}
P_{\Delta T}({\bf k}) = \tilde{T}_b^2 \,  \left[\bar{x}_H^2 \,
P_{\delta \delta} +  P_{x x} - 2 \bar{x}_H \, P_{x \delta} + P_{x
\delta x \delta} \right].
\label{eq:PTb}
\end{eqnarray}
We construct $P_{\delta \delta}$ with the halo model \citep{cooray02} and $P_{xx}$ and $P_{x \delta}$ with the methods described in \citet{mcquinn05}.  The fourth moment $P_{x \delta x \delta}$ follows from $P_{\delta \delta}, \, P_{xx}$, and $P_{x \delta}$ using the hierarchical ansatz.

\begin{figure}
\begin{center}
\resizebox{8cm}{!}{\includegraphics{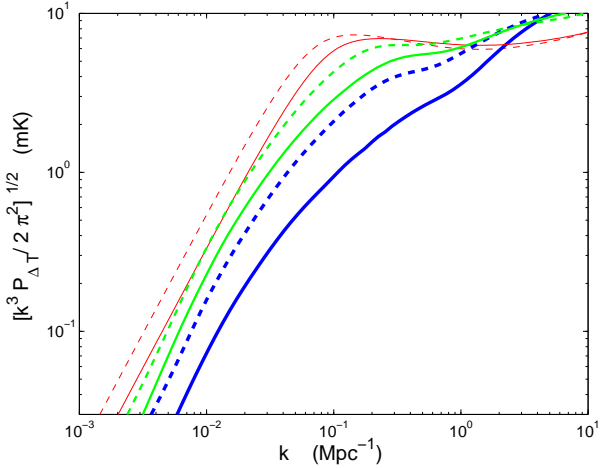}}\\%
\end{center}
%\plotone{zetam.eps}
\caption{21 cm brightness temperature power spectra for $\bxio = 0.25,\,0.4,\, {\rm and}\, 0.7$ (thick, medium, and thin curves, respectively).  The solid and dashed curves take $\zeta \propto m^0$
and $\zeta \propto m^{2/3}$, respectively. }
\label{fig:21cm}
\end{figure}

In Figure~\ref{fig:21cm}, we plot $[k^3 \, P_{\Delta T}(k)/2
\pi^2]^{1/2}$ at $z = 12$ for the original FZH04 model in which $\zeta
\propto m^0$ (solid curves) and for a modified model in which $\zeta
\propto m^{2/3}$ (dashed curves).  For each model, we show $\bar{x}_i
= 0.25,\,0.4$, and $0.7$ (thick, medium, and thin curves,
respectively).  We see clear differences in the power spectra: both
the peak (measuring $R_c$) and the amplitude differ between the
models.  The disparity decreases as a function of $\bar{x}_i$,
differing by about a factor of two for scales $k < 1$ Mpc at
$\bar{x}_i = 0.25$.  This is to be expected from
Figure~\ref{fig:dndr}\emph{b}: the differences between the bubble size
distributions decrease with increasing ionized fraction, primarily
because of the shape of the matter density fluctuations.  While we
have shown results for $z=12$, the curves depend primarily on the
bubble size distribution and show similar differences at other
redshifts (see Fig.~\ref{fig:dndr}\emph{a}).

%%%%%%%%%FIGURE 8: 21cm signal

\citet{bowman05} find that the Mileura Widefield Array (MWA) will be most sensitive to wavenumbers
$0.01 \la k \la 0.1 \Mpc^{-1}$ (This is approximately true for most of the other planned telescopes as well.).  These scales are larger than the effective bubble sizes for most $\bxio$ in our model.  At such scales, equation (\ref{eq:PTb}) approaches
\begin{equation}
P_{\Delta T_b}({\bf k}) \rightarrow \tilde{T}_b^2 ~ \left[\bar{x}_H^2 + b^2
-2 \bar{x}_H b \right] \, P_{\delta \delta},
\label{eq:Plimit}
\end{equation}
where $b$ is the volume-averaged linear bias of the bubbles
\citep{mcquinn05}.  In the $\zeta \propto m^{2/3}$ model, the bubbles
are more highly biased.  Intuitively, this is because they trace the
most massive galaxies, which are of course also the most biased.
Hence, for a fixed $\bar{x}_i$, the large-scale signal increases in
the $\zeta \propto m^{2/3}$ model.  Conversely, in the small scale
limit $ k \ga 1 ~ \Mpc$, density fluctuations in the
remaining neutral gas dominate the signal.  It has the limiting form
$P_{\Delta T_b}({\bf k}) \rightarrow \tilde{T}_b^2 \, \bxh^2 \,
P_{\delta \delta}$.  We indeed see that the curves with the same
$\bxio$ converge at small scales; this regime contains little
information about the ionizing sources.

More generally, Figure \ref{fig:21cm} demonstrates how sensitive the
21 cm signal is to the bubble size distribution.  These observations
will thus provide powerful constraints on $n_b(m)$.  They should also
be able to probe stochastic fluctuations in the galaxy population (\S
\ref{flucs}), which will decorrelate the $x_i$ and the density fields.
Other, more subtle properties (such as the halo mass function, \S
\ref{nh}) may even be discernable in the distant future.  However, we
must note that the detailed power spectra will depend on other effects
that we have not included in this simple analytic treatment -- most
important, we have assumed spherical bubbles throughout.  These
effects can be tested with numerical realisations of the models
described here (such as those in \citealt{zahn05}) or with full
cosmological simulations that also include radiative transfer;
preliminary tests indicate that our analytic model roughly matches the
numerical results \citep{mcquinn05}, except when $\bar{x}_i > .85$ (at which point our picture of isolated bubbles breaks down).

\subsection{The kSZ effect}
\label{ksz}
 
The peculiar velocities of the \hii regions imply that the scattering of CMB photons off free electrons in the bubbles will impart a red- or blueshift to the scattered photons, creating hot and cold spots in the CMB \citep{aghanim96}.  The power spectrum of temperature anisotropies produced by such scatterings, part of a broader class of anisotropies termed the kinetic Sunyaev Zel'dovich (kSZ) effect  \citep{sunyaev80}, have been calculated for several analytic models of reionization (e.g. \citealt{gruzinov98, knox98, santos03, zahn05}).  Recently, \citet{mcquinn05} used the FZH04 model to demonstrate that  the kSZ anisotropies are quite sensitive to both the duration of reionization and the bubble size distribution.  

In the original FZH04 model, ${\it l}^2 C^{\rm kSZ}_{\it l}$ peaks at approximately the angular scale corresponding to $R_c$ when $\bxio \approx 0.5$ (around ${\it l} = 3000$) and has a substantial width in ${\it l}$ owing to the distribution of bubble sizes.  For ${\it l} \ga 5000$, the primordial component to the
power spectrum becomes negligible owing to Silk damping, providing a window in which we can observe these secondary anisotropies.  The kSZ signal lies well below another secondary anisotropy, the thermal Sunyaev-Zel'dovich (tSZ) effect, at all angular scales.  Fortunately, the tSZ anisotropies have a unique frequency dependence, which should allow them to be cleanly removed.  The Atacama Cosmology Telescope (ACT) and the South Pole telescope (SPT) are designed to detect the kSZ anisotropies and should be operational as early as 2007.\footnote{For more information on these experiments, see http://www.hep.upenn.edu/act/act.html and http://astro.uchicago.edu/spt.}  We first summarise our formalism for calculating the kSZ signal (see \citealt{mcquinn05} for more details) and then discuss how ${\it l}^2 C^{\rm kSZ}_{\it l}$ changes if we allow $\zeta \propto m^\alpha$.

Thomson scattering of CMB photons off free electrons with a bulk peculiar velocity produces a temperature anisotropy along the line of sight ${\bf \hat{n}}$:
\begin{equation}
{\frac{\Delta T}{T}({\bf \hat{n}})}_{\rm kSZ} = \sigma_T
    \; \int{ \deriv \eta \; e^{-\tau(\eta)} \; a(\eta) \, n_e(\eta, {\bf \hat{n}}) \;
({\bf \hat{n}} \cdot {\bf v}  }),
\label{anisotropy}
\end{equation}
where $a$ is the scale factor, $\tau (\eta)$ is the Thomson optical depth to the scatterer at conformal time $\eta$, ${\bf v}({\bf \hat{n}}, \eta) $ is the peculiar velocity of the scatterer, and the electron number density is
\begin{equation}
n_e(\eta, {\bf \hat{n}}) = \bar{n}_e (\eta) \; \bar{x}_i(\eta) \; [1
+ \delta(\eta, {\bf \hat{n}})+ \delta_{x}(\eta, {\bf \hat{n}})].
\label{eq:ne}
\end{equation}
Here, $\bar{n}_e$ is the average number density of electrons (free and bound).\footnote{Note that eq. (\ref{eq:ne}) is not a formally rigorous perturbation expansion. However, it suffices for our purposes because the second order $\delta_x \delta$ term contributes a negligible fraction of the total kSZ signal. }

One might expect the dominant contribution to the kSZ angular power spectrum to come at zeroth order in the perturbations $\{\delta,\, \delta_x\}$, arising only from the velocity field.  However, because only the component of the velocity along the line-of-sight contributes to the anisotropy and because it enters as an integral along the line-of-sight (eq.~\ref{anisotropy}), the signal averages over many troughs and peaks, making the net contribution effectively zero for the modes of interest. The lowest order
contributions arise from modes perpendicular to the line-of-sight, which occur at second order in $\{\delta, \delta_x\}$.  \citet{kaiser92} and \citet{jaffe98} showed that at small scales and in the flat sky approximation the angular power spectrum of $(\Delta T/T)_{\rm kSZ}$ can be written as
\begin{equation}
    C^{\rm kSZ}_{\it l} =  \int{\frac{d\eta}{x^2} \;
    {W(\eta)}^2 \, P_{\rm kSZ}({\it l}/x, \eta)},
    \label{cl}
\end{equation}
where $x = \eta_0 - \eta$, $W =\sigma_T \,\bar{n}_e(\eta_0) \, a^{-1} \, \bar{x}_i^2 \, e^{-\tau}$, and \citep{ma02, santos03}
\bqa
P_{\rm kSZ}({\bf k})  & =  & \frac{1}{2}
\int \frac{\deriv^3{\bf k'}}{(2 \pi)^3}  \left\{ \left(1 - {\hat{\bf k}\cdot \hat{\bf k}'}^2 \right)
P_{vv}(k') \right. \nonumber \\ 
& & \times \left. \left[ P_{x x}(|{\bf k} - {\bf k}'|) + 2 P_{x \delta}(|{\bf k} - {\bf k}'|) \right] \right\}.
\label{eq:Pksz} 
\eqa We have omitted the subdominant terms in equation
(\ref{eq:Pksz}), and we use linear theory to calculate the power
spectrum of velocity fluctuations $P_{vv}$ \citep{mcquinn05}.

Equation (\ref{eq:Pksz}) also omits a $P_{\delta \delta}$ term that
enters the sum within the brackets.  This contribution is called the
Ostriker-Vishniac (OV) effect; it has been studied in much more depth
than the other ``patchy'' reionization terms (e.g.,
\citealt{ostriker86, jaffe98, hu00}), because it relies on much
simpler physics.  However, the other two terms dominate during the
reionization epoch, so we do not include it in our analysis.  On the
other hand, the OV effect continues growing past reionization, and for
reasonable reionization histories in the FZH04 model, the total OV
signal tends to be $\sim 1$--$5$ times larger than the patchy
contributions around ${\it l} = 5000$ \citep{zahn05, mcquinn05}.
Fortunately, simulations should be able to calculate the OV
anisotropies to reasonable precision, given a cosmology and a
reionization history (they simply trace $P_{\delta \delta}$ for the
baryons).  In that case it can be isolated from the patchy
reionization terms in which we are most interested.

\begin{figure}
\begin{center}
\resizebox{8cm}{!}{\includegraphics{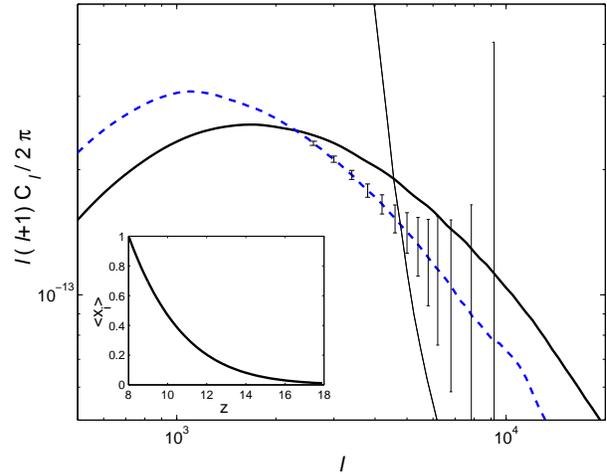}}\\%
\end{center}
%\plotone{zetam_kSZ.eps}
\caption{Angular power spectrum of the kSZ signal for models with $\zeta \propto m^0$ (solid curve) and $\zeta \propto m^{2/3}$ (dashed curve).  The inset shows the assumed ionization history.  The thin solid line shows the primordial anisotropies, which dominate when ${\it l} \la 5000$.  The error bars show the 1-$\sigma$ errors for the ACT experiment. }
\label{fig:kSZ}
\end{figure}

Figure \ref{fig:kSZ} shows the kSZ angular power spectrum for models
with $\zeta \propto m^0$ (solid curve) and with $\zeta \propto
m^{2/3}$ (dashed curve).  Both assume the same $\bxio(z)$ (see inset).
In the $\zeta \propto m^{2/3}$ model, the bubbles are larger
throughout reionization.  This is reflected in the power spectrum: it
peaks at larger scales (${\it l} = 1000$ rather than ${\it l} = 2000$)
and is suppressed by $\sim 30\%$ relative to the constant $\zeta$
model on small scales (${\it l} \ga 3000$).  This occurs because fewer
small bubbles appear.  Unfortunately, the large scales on which the
peak occurs will be washed out by the primordial anisotropies (the
thin line in Fig. \ref{fig:kSZ}).\footnote{However, note that the kSZ
effect on scales ${\it l} \sim 1000$--$5000$ will bias cosmological
parameter measurements for future missions such as the Planck
satellite.  Thus the large scale behaviour is still detectable; it is
just more difficult to isolate owing to additional degeneracies
it suffers with the cosmological parameters that determine the primordial
anisotropies \citep{santos03, zahn05}.}  In Figure~\ref{fig:kSZ}, we
plot the 1-$\sigma$ ACT error bars (assuming perfect removal of
foregrounds as well as of the tSZ and OV anisotropies).  If these
contaminants can be adequately removed, the ACT should be able to
distinguish these two models of reionization through the amplitude of
the fluctuations on relatively small scales.

%%%%%%%%%FIGURE 9: kSZ

We found that the bubble size distribution figures prominently in the
angular distribution of power of the kSZ signal.  However, the
duration of reionization is primarily responsible for determining the
amplitude of the kSZ signal \citep{mcquinn05}.  The history we have
chosen is relatively narrow and has $\tau = 0.095$, which is about
2-$\sigma$ below the WMAP best fit value of $\tau= 0.17 \pm 0.04$
\citep{kogut03}.  The relative differences between the two models
should be similar regardless of the redshift of reionization or its
duration, because they depend primarily on $n_b(m)$, which is nearly
independent of redshift for a constant $\bxio$.  However, the amplitude of the signal
could be up to a factor of five larger if reionization is
extended, which would make these differences even easier to measure.
The most important caveat is that the shape of the power spectrum on
scales where the primordial anisotropies are negligible changes
relatively little between the $\zeta \propto m^0$ and $m^{2/3}$
models.  Distinguishing the size distribution from, for example, a
slightly faster reionization history, may be difficult and will
require considerably more detailed modeling with simulations (c.f.,
Fig.~5 of \citealt{mcquinn05}).

\section{Discussion}
\label{disc}

In this paper, we have taken a closer look at the key properties of
the FZH04 analytic model for the growth of \hii regions during
reionization.  First, we examined the origin of the characteristic
scale $R_c$.  We showed that it has a simple physical interpretation
as the scale on which a ``typical" density fluctuation is large enough
to ionize itself.  The scale at which this condition is fulfilled
depends on the bias of the underlying galaxy population, which gives a
boost to the matter density fluctuations.  Increasing the mean
bias drives reionization to larger scales by allowing weaker density
fluctuations (which appear on larger scales) to ionize themselves.

Another key feature of the FZH04 model is that the bubble size
distribution narrows throughout reionization while $R_c$ becomes less
and less sensitive to the input parameters.  We showed that this
primarily results from the shape of the underlying matter power spectrum: at
Mpc scales, the effective power law slope is $n \sim -2.3$, while for
$R \ga 30 \Mpc$, $n \sim -1$.  In the former case, density
fluctuations collapse at a similar time over a wide range of scales,
making $R_c$ quite sensitive to the input parameters.  For $n \sim
-1$, on the other hand, the scale dependence is much weaker.

We then considered two modifications to the FZH04 formalism.  One is
the underlying halo mass function: FZH04 use the \citet{press74}
distribution, which differs in detail from the mass function observed
in simulations of the low-redshift universe \citep{jenkins01}.  We
showed that the \citet{sheth99} parameterisation, which better matches
simulations, barely alters the bubble mass function.  The reason
is that the two halo mass functions have similar dependence on the
underlying density field.  We also allowed the ionizing efficiency
parameter $\zeta$ to be a function of galaxy mass.  This may be
appropriate if, for example, feedback affected the star formation
efficiency, IMF, or escape fraction inside galaxies.  We showed that
weighting massive galaxies more heavily increases $R_c$ because such
galaxies are more highly biased.  Under reasonable assumptions, this
could increase $R_c$ by several times throughout the early and middle
stages of reionization.

This difference in the bubble sizes could be directly observable in
the near future.  In particular, if massive galaxies drive
reionization the 21 cm brightness temperature power spectrum increases
by about a factor of two early in reionization on scales accessible to
the next generation of radio telescopes.  The shape of the induced CMB
angular power spectrum also changes at small scales, although it
may be more difficult to extract information about the sources from
this integrated measure because of degeneracies with the duration of
reionization.  Nevertheless, we have shown explicitly that both of
these techniques offer the exciting possibility of learning about the
sources driving reionization -- which may remain hidden from us for
many years -- through measurements of the IGM on $\sim 10 \Mpc$
scales, which may be more accessible.  Another possibility is to
constrain the \hii distribution with the observed abundance and clustering of
Ly$\alpha$-selected galaxies \citep{furl04-lya, furl05-lyagal}.

Finally, we examined the conditions under which stochastic
fluctuations in the galaxy distribution can significantly modify the
bubble size distribution.  Such variations are inevitable because
galaxies do not trace the dark matter distribution precisely.  Using a
simple Poisson model (which is approximately correct at lower
redshifts; \citealt{casas02}), we showed that such fluctuations are
typically $\la 10\%$ once $R_c \ga 3 \Mpc$.  In this case, they should
have only a modest effect on the bubble sizes, except during the early
stages of reionization.  However, the fluctuations could be much more
important if massive galaxies are responsible for reionization,
because then the effective number density of sources falls
dramatically.  In such a scenario, numerical simulations of the full
reionization process may be necessary to predict the bubble
characteristics.

One particularly interesting application of these results is the possibility of reionization by quasars.  The bright quasars detected by the Sloan Digital Sky Survey do not seem able to reionize the universe at $z \sim 6$, provided that the luminosity function is not extremely steep \citep{fan04}.  However, it remains possible that a population of relatively faint quasars could have made significant contributions (though see \citealt{dijkstra04}).  Our models suggest that such a picture would have a number of implications for the \hii regions during reionization.  First, if quasars preferentially traced massive galaxies, they would have produced characteristically large bubbles.  This may be particularly relevant if, as recently suggested by \citet{lidz05}, faint quasars (as well as bright ones) populated massive galaxies, or if quasars grew most efficiently in massive haloes \citep{wyithe02}.  Moreover, if quasars only populated massive galaxies, their stochastic fluctuations would be larger.  This is particularly relevant because they also likely had short lifetimes with highly variable luminosities \citep{hopkins05b, hopkins05}; Figure~\ref{fig:recz8} shows that the resulting emissivity fluctuations would cause the bubbles to grow in spurts.  These global signatures of quasar reionization complement the local signature proposed by \citet{zaroubi05}:  the profile of ionization fronts could help distinguish hard and soft sources.  

We have found that the major properties of the FZH04 model of
reionization are robust to several of the simplifications inherent to
it.  Most important, large-scale ($\ga 10 \Mpc$) inhomogeneities in
the ionized fraction seem unavoidable.  This is crucial to, e.g., 21
cm measurements of reionization, because the next generation of
telescopes will only be sensitive to structures several Mpc across.
The existence of a characteristic scale is also generic to any model
in which galaxies drive reionization.  On the other hand, the FZH04
model still contains a number of simplifications -- crucially, the
assumption of isolated spherical bubbles -- that may be particularly
problematic during the late stages of reionization.  Cosmological
simulations will still be needed to predict detailed observational
signatures as well as to test the behaviour of our analytic models.
Such tests are challenging because of the wide separation in scales
between the galaxies and the \hii regions.  One promising hybrid
approach is to implement the FZH04 model in numerical simulations or
even simple gaussian random fields \citep{zahn05}.  \citet{mcquinn05}
showed that the kSZ power spectrum from our fully analytic model and
from numerical realisations match each other rather well, which bodes
well for the utility of these simple analytic models.

We thank M. Kamionkowski for helpful comments.  This work was supported by NSF grants ACI 96-19019, AST 00-71019, AST 02-06299, and AST 03-07690, and NASA ATP grants NAG5-12140, NAG5-13292, and NAG5-13381.

%\bibliographystyle{mnras}
%\bibliography{Ref_alph}

\end{document}